\documentclass[12pt,preprint]{emulateapj}

\shorttitle{X-rays from Millisecond Pulsars in 47 Tuc}
\shortauthors{Bogdanov et al.}

\begin{document}

\title{Chandra X-ray Observations of Nineteen Millisecond Pulsars \\ in the Globular Cluster 47 Tucanae}

\author{Slavko Bogdanov\altaffilmark{1}, Jonathan E. Grindlay\altaffilmark{1}, Craig O. Heinke\altaffilmark{1,2,3}, \\ Fernando Camilo\altaffilmark{4}, Paulo C. C. Freire\altaffilmark{5}, and Werner Becker\altaffilmark{6}}
\altaffiltext{1}{Harvard-Smithsonian Center for Astrophysics, 60 Garden St., Cambridge, MA 02138; sbogdanov@cfa.harvard.edu, josh@cfa.harvard.edu, cheinke@cfa.harvard.edu}

\altaffiltext{2}{Northwestern University, Department of Physics and
Astronomy, 2145 Sheridan Rd., Evanston IL 60208; cheinke@northwestern.edu} 

\altaffiltext{3}{Lindheimer Postdoctoral Fellow}

\altaffiltext{4}{Columbia Astrophysics Laboratory, Columbia University, 550 West 120th Street, New York, NY 10027; fernando@astro.columbia.edu}

\altaffiltext{5}{National Astronomy and Ionosphere
Center, Arecibo Observatory, HC3 Box 53995, PR 00612;
pfreire@naic.edu}

\altaffiltext{6}{Max-Planck-Institut f\"ur Extraterrestrische Physik, D-85740 Garching bei M\"unchen, Germany; web@mpe.mpg.de}

\begin{abstract}
We present spectral and long-timescale variability analyses of
\textit{Chandra X-ray Observatory} ACIS-S observations of the 19
millisecond pulsars (MSPs) with precisely known positions in the
globular cluster 47 Tucanae. The X-ray emission of the majority of
these MSPs is well described by a thermal (blackbody or neutron star
hydrogen atmosphere) spectrum with a temperature $T_{\rm
eff}\sim(1-3)\times10^6$ K, emission radius $R_{\rm eff}\sim0.1-3$ km,
and luminosity $L_{X}\sim10^{30-31}$ ergs s$^{-1}$. For several MSPs,
there is indication that a second thermal component is required,
similar to what is seen in some nearby field MSPs. The observed
radiation most likely originates from the heated magnetic polar caps
of the MSPs.  The small apparent scatter in $L_{X}$ is consistent with
thermal emission from the polar caps of a global dipole field although
the small emission areas may imply either a more complex small-scale
magnetic field configuration near the neutron star surface or
non-uniform polar cap heating.  The radio-eclipsing binary MSPs 47 Tuc
J, O, and W show a significant non-thermal (power-law) component, with
spectral photon index $\Gamma\sim 1-1.5$, which most likely originates
in an intrabinary shock formed due to interaction between the
relativistic pulsar wind and matter from the stellar companion.  We
re-examine the X-ray--spindown luminosity relation ($L_{X}-\dot{E}$)
and find that for the MSPs with thermal spectra
$L_{X}\propto\dot{E}^{\beta}$, where $\beta\sim0.2\pm1.1$. Due to the
large uncertainties in both parameters the result is consistent with
both the linear $L_{X}-\dot{E}$ relation and the flatter
$L_X\propto\dot{E}^{0.5}$ predicted by polar cap heating models.  In
terms of X-ray properties, we find no clear systematic differences
between MSPs in globular clusters and in the field of the Galaxy.  We
discuss the implications of these results of the present understanding
of the X-ray emission properties of MSPs.
\end{abstract}


\section{INTRODUCTION}
Millisecond pulsars (MSPs) represent a distinct population of
rotation-powered pulsars, characterized by short spin periods,
$P\lesssim25$ ms, and small intrinsic spin-down rates, $\dot{P}_{i}\sim
10^{-20}$, implying relatively low surface magnetic dipole field
strengths $B_{\rm surf}\propto(P\dot{P}_{i})^{1/2}\sim10^{8-10}$ G and
large characteristic spindown ages $\tau\equiv P/2\dot{P}_{i}\gtrsim
1$ Gyr.  These objects have been studied extensively at radio
wavelengths since their discovery \citep{Back82} but at X-ray
energies, where the bulk of observable radiation is expected, this has
become possible only relatively recently, owing to their intrinsic
faintness ($L_{X}<10^{33}$ ergs s$^{-1}$).  The \textit{ROSAT}
observatory was the first to detect X-ray emission from MSPs,
including 10 in the field of the Galaxy and one in the globular
cluster M28 \citep[see][and references therein]{Beck99}. In recent
years, the unprecedented sensitivity and spatial resolution of the
\textit{Chandra X-ray Observatory} have allowed the detection of X-ray
counterparts to radio MSPs in several globular clusters: 47 Tucanae
and NGC 6397 \citep{Grind02}, NGC 6752 \citep{DAmico02}, M4
\citep{Bass04b}, possibly M30 \citep{Ran04}, and M71 (Elsner et al.,
in preparation).  In addition, the \textit{Chandra} and
\textit{XMM-Newton} telescopes have allowed much more detailed studies
of the existing sample of X-ray-detected MSPs than previously possible
\citep*{Zavlin02,Beck02,Beck03,Webb04a,Webb04b}.

\begin{figure*}[t!]
\begin{center}
\includegraphics[width=0.90\textwidth]{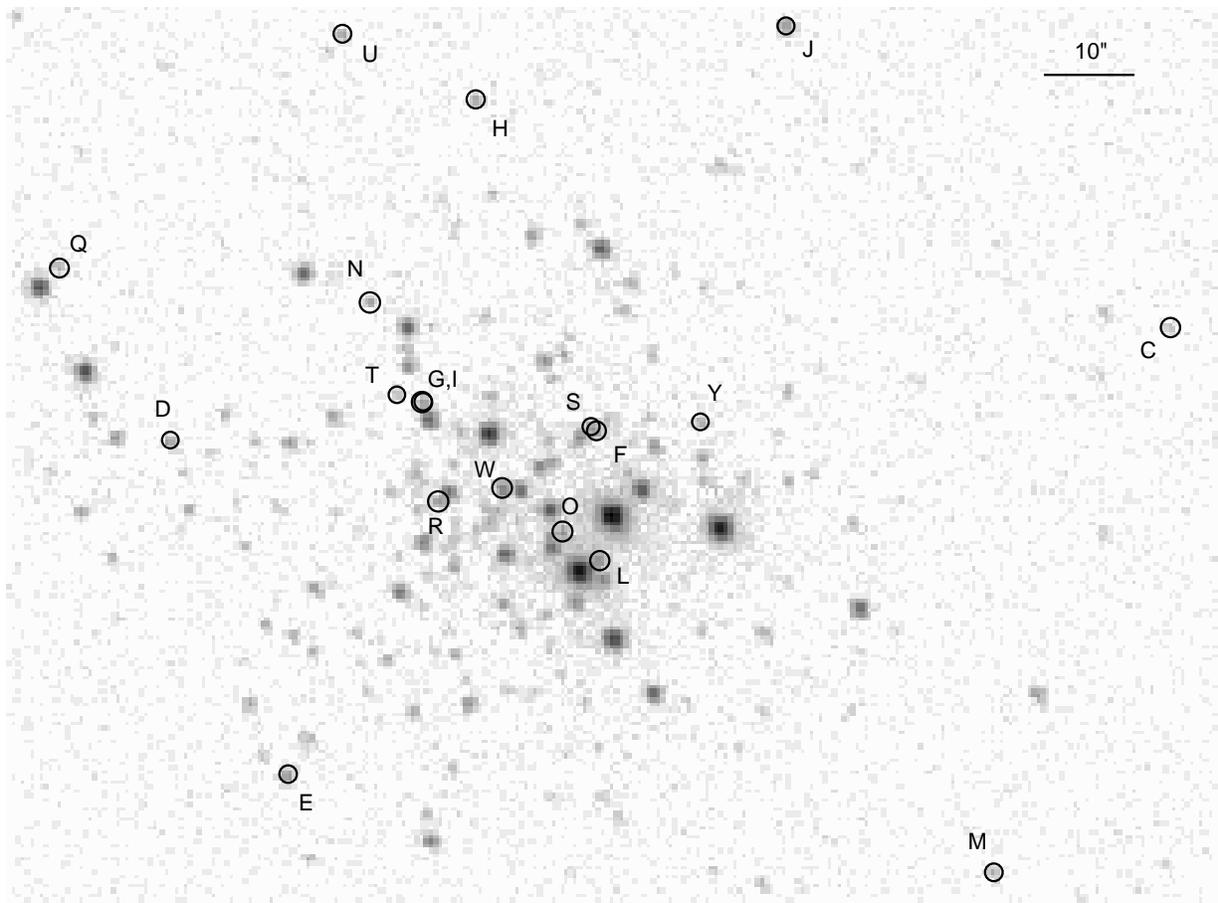}
\caption{A merged $0.3-6$ keV image of all four \textit{Chandra}
ACIS-S observations of 47 Tucanae showing the positions of the X-ray
counterparts of 19 millisecond pulsars. The $1''$ circles are centered
on the radio MSP locations and enclose $\ge90$\% of the total energy
for $0.5-1$ keV thermal sources. The grayscale corresponds to number
of counts increasing logarithmically from zero (\textit{white}) to 8306
(\textit{black}). North is up and east is to the left.}
\end{center}
\end{figure*}

The X-ray emission detected from MSPs is observed to be both of
non-thermal and thermal character. In the X-ray bright MSPs, B1821--24
and B1937+21, the narrow X-ray pulse profiles indicate the emission is
beamed and, therefore, must be non-thermal radiation originating in
the pulsar magnetosphere \citep{Beck99}.  Alternatively, as seen in
the ``black widow'' pulsar, B1957+20, non-pulsed non-thermal X-rays
can be produced by interaction of the relativistic particle wind from
the pulsar with matter from a stellar companion or the interstellar
medium \citep{Stap03}. Thermal emission observed from nearby MSPs
\citep[see e.g.][]{Zavlin02, Beck02}, is believed to originate from
the magnetic polar caps of the underlying neutron star (NS), heated by
a backflow of energetic particles from the pulsar magnetosphere
\citep[see e.g.][and references therein]{Hard02a,Zhang03}.

The detection of 16 MSPs in the globular cluster 47 Tucanae (NGC 104,
henceforth 47 Tuc) by the \textit{Chandra} ACIS-I instrument has
granted an excellent opportunity for a detailed study of the physical
properties of this class of objects \citep{Grind01, Grind02}.  These
pulsars were discovered at radio wavelengths by \citet{Man90,Man91},
\citet{Rob95}, and \citet{Camilo00}. With further pulse timing
observations, \citet{Freire01a,Freire03} were able to obtain precise
positions for 16 of the 22 MSPs presently known in this globular
cluster \citep{Camilo05}. In addition, \citet{Edm02} detected the
optical counterpart of 47 Tuc W, which does not have an accurate radio
timing position, by matching the binary phase and period of an optical
candidate with that of the radio pulsar. Finally, timing solutions
have recently been obtained for 47 Tuc R and Y (Freire et al., in
preparation) thus increasing the number of MSPs with known positions
in this cluster to 19. \citet{Grind02} have reported in detail on the
analysis of the X-ray emission properties of the 47 Tuc MSPs, using
the \textit{Chandra} ACIS-I observations mentioned above.  However,
due to the inherent faintness of these sources ($L_{X}\sim10^{30-31}$
ergs s$^{-1}$) and the limited exposure time (72 kiloseconds of total
data obtained resulting in 1-30 detected counts for each MSP), the
exact nature of the X-ray emission could not be determined with great
confidence, although some interesting constraints could be placed.  In
the meantime, additional X-ray observations of 47 Tuc have been
carried out, allowing further study of the properties of globular
cluster MSPs as well as MSPs in general.

In this paper, we present spectral and long-timescale variability
analyses of the X-ray counterparts of MSPs in the globular cluster 47
Tuc based on deep \textit{Chandra} ACIS-S observations. This paper
extends the work presented in \citet{Heinke05}. The work is organized
in the following manner: \S 2 describes the observations and data
reduction procedure; in \S 3 we discuss the X-ray spectral properties
of the MSPs, while in \S 4 we investigate the long-term temporal
behavior of the X-ray emission; in \S 5 we compare the X-ray and radio
pulsar properties. Finally, in \S 6 we discuss the implications of the
results presented and offer conclusions in \S 7.

\section{OBSERVATIONS AND DATA REDUCTION}

The data set presented here consists of four 65-kilosecond
observations of the core of the globular cluster 47 Tuc with the S3
chip of the \textit{Chandra} ACIS-S detector at the focus.  The first
three observations were performed between 2002 September 29 and
October 3, while the fourth was carried out on 2002 October 11. These
observations have resulted in the detection of $\sim$300 X-ray sources
within the half-mass radius of 47 Tuc, including X-ray counterparts to
all of the 19 MSPs with known positions.  In this present paper we
focus on the counterparts of the radio MSPs; for a comprehensive
analysis of the other X-ray sources, see \citet{Heinke05}. The initial
data reduction and image processing were performed using the
CIAO\footnote{Chandra Interactive Analysis of Observations
(http://asc.harvard.edu/ciao/)} 3.0 software package and are also
described in detail by \citet{Heinke05}. Figure 1 shows the combined
image of all four observations with $1''$ circles centered at the
radio position of each MSP.  The CIAO tool WAVDETECT was able to
detect an X-ray source at the position of each of the radio MSPs,
including 47 Tuc C, L, M, Q, and T, which were not detected or only
marginally detected in the 2000 ACIS-I observations
\citep{Grind02}. The MSPs 47 Tuc G and I, separated by just
$0\farcs12$, could not be individually resolved, whereas 47 Tuc F and
S, separated by $0\farcs7$, were again identified as a single
elongated source. In this paper, we also include analyses of MSP W,
whose optical counterpart was found by \citet{Edm02} to coincide with
the X-ray source W29 \citep{Grind01}, and the MSPs R and Y, whose
recently obtained radio timing solutions (Freire et al., in
preparation) place them at the positions of the X-ray sources W198 and
W82, respectively.

To facilitate the spectral and variability analyses, the counts for
each MSP in the 0.3--8 keV energy band were extracted from circular
regions with $1''$ radii around the radio MSP positions. These circles
enclose $\ge$$90$\% of the total energy for 0.5--1 keV thermal
sources.  A slightly more involved extraction method was required for
47 Tuc F and S, whose extraction circles overlap, as evident in Figure
1. The relative contribution of counts from each MSP was determined by
maximum likelihood fits of two-dimensional Gaussians to the five X-ray
sources in the vicinity of the radio positions of F and S using the
\textit{Sherpa} application in CIAO.  This procedure was carried out
in the 0.3--4 keV range, as well as in three narrow bands (0.3--0.8
keV, 0.8--1.5 keV, and 1.5--4 keV) to allow an approximate spectral
analysis. Using this method we find that $62\pm7$\% of the total
counts can be attributed to S.  For 47 Tuc G and I, this was not
possible as their radio positions are only $0\farcs12$ apart. However,
the count rate observed from their position is roughly twice that of a
typical MSP in the sample and is, thus, consistent with being combined
emisson from both MSPs. The net counts for most MSPs were obtained by
subtracting the background rate, obtained from three source-free
regions on the image, from the total.

In general, the background noise in the energy range of interest was
found to be very low ($\sim$1 background count per pixel over
$\sim$260 ks of total exposure time). This was not the case for 47 Tuc
L, which is located very near the bright X-ray source X9
\citep{Grind01}. The background for this MSP was obtained by
considering a section of an annulus of width $2''$ around the
positions of X9, at the distance to L, while excluding the $1''$
extraction circle around the MSP. Using this background region we find
that $\sim$30\% of the gross number of counts are due to X9.  For 47
Tuc R, situated $1\farcs3$ from the bright source W24, this
procedure was not possible due to a combination of the smaller
distance to W24 and the large pixel size of the ACIS
detector. However, the distribution of counts within the $1''$
extraction region of R suggests that W24 contributes with no more than
10\% to the total counts. In the case of MSP O, the degree of
contamination is difficult to gauge due to severe crowding. In
addition, the X-ray counterpart of 47 Tuc O, W39, is very likely a
blend of the MSP and a variable source (see \S 4).

\begin{figure}[t!]
\begin{center}
\includegraphics[width=0.48\textwidth]{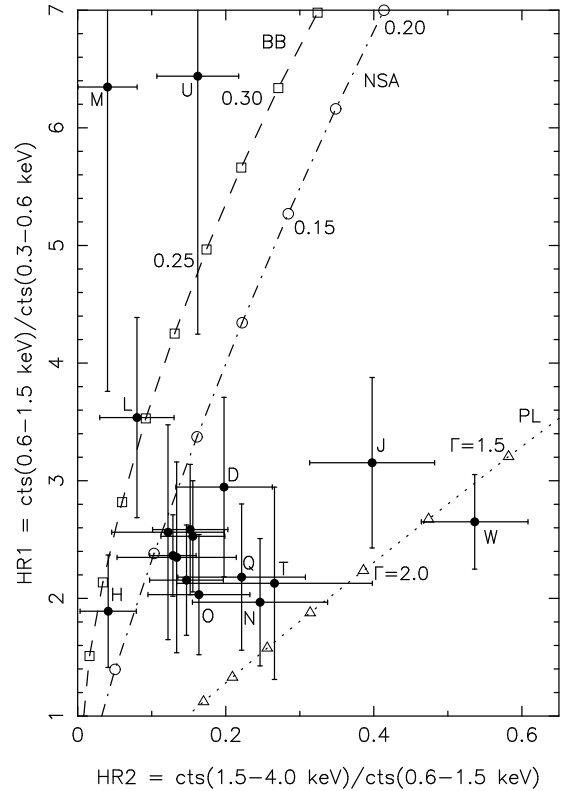}
\caption{X-ray color-color diagram for the MSPs in 47 Tuc. The lines
represent model tracks for absorbed power-law (\textit{dotted}),
blackbody (\textit{dashed}), and neutron star hydrogen atmosphere
(\textit{dot-dashed}) spectra convolved with the ACIS-S instrument
response. The open symbols mark values for the model tracks in
intervals of 0.025 keV for the blackbody (\textit{squares}) and
hydrogen atmosphere (\textit{circles}) temperatures and 0.25 for the
power-law photon index (\textit{triangles}).}
\end{center}
\end{figure}

\begin{deluxetable}{lcccc}
\tabletypesize{\small} 
\tablecolumns{5} 
\tablewidth{0pc}
\tablecaption{Net counts from the 19 MSPs in 47 Tuc.}
\tablehead{ \colhead{ } &
\colhead{Soft} & \colhead{Medium} & \colhead{Hard} & \colhead{Total Cnts}\\
\colhead{MSP} & \colhead{(0.3-0.6 keV)} & \colhead{(0.6-1.5 keV)} &
\colhead{(1.5-4 keV)} & (0.3-4 keV)}
\startdata
 C & $12 \pm 4$ & $29 \pm 5$ & $4 \pm 2$ & $45 \pm 7 $ \\
 D & $20 \pm 5$ & $60 \pm 8$ & $12 \pm 4$ & $92 \pm 10$ \\
 E & $39 \pm 6$ & $74 \pm 9$ & $11 \pm 3$ &  $124 \pm 11$ \\
 F+S\tablenotemark{a} & $67 \pm 8$ & $159 \pm 13$ & $20 \pm 5$ & $246 \pm 16$ \\
 F\tablenotemark{a} & $15 \pm 12$ & $83 \pm 21$ & $12 \pm 7$  & $93 \pm 18 $ \\
 G+I & $40 \pm 6$ & $102 \pm 10$ & $16 \pm 4$  & $158 \pm 12$ \\
 H & $24 \pm 5$ & $46 \pm 7$ & $2 \pm 2$ & $72 \pm 9$ \\
 J & $25 \pm 5$ & $80 \pm 9$ & $32 \pm 6$ & $137 \pm 12$ \\
 L & $37 \pm 8$ & $132 \pm 14$ & $11 \pm 7$ & $180 \pm 18$ \\
 M & $7  \pm 3$ & $47 \pm 7$ & $2 \pm 2$ &  $56 \pm 8$ \\
 N & $20 \pm 5$ & $40 \pm 7$ & $10 \pm 3$ & $70 \pm 9$  \\
 O & $30 \pm 6$ & $61 \pm 9$ & $10 \pm 4$ & $101 \pm 12$  \\
 Q & $18 \pm 4$ & $40 \pm 6$ & $9 \pm 3$ &  $67\pm 8$ \\
 R & $41 \pm 8$ & $105 \pm 11$ & $16 \pm 5$ & $162 \pm 14$  \\
 S\tablenotemark{a} & $52 \pm 12$ & $76 \pm 21$ & $8 \pm 7$  &  $153 \pm 20$ \\
 T & $10 \pm 3$ & $22 \pm 5$ & $6 \pm 3$  & $38 \pm 7$ \\
 U & $10 \pm 3$ & $67 \pm 8$ & $11 \pm 3$  & $88 \pm 9$ \\
 W & $60 \pm 8$ & $160 \pm 13$ & $86 \pm 9$ & $306 \pm 18$ \\
 Y & $16 \pm 5$ & $41 \pm 7$ & $5\pm 3$ & $62 \pm 9$ \\
\enddata
\tablenotetext{a}{For F and S, we
quote the combined counts as well as the counts for each MSP
apportioned using the procedure described in the text.}

\end{deluxetable}

\section{SPECTRAL ANALYSIS}

To get a general sense of the X-ray properties of the MSPs, we have
computed the hardness ratios HR1=(medium counts)/(soft counts) and
HR2=(hard counts)/(medium counts) for each object and constructed an
X-ray color-color diagram. The soft (0.3--0.6 keV), medium (0.6--1.5
keV), and hard (1.5--4 keV) energy bands were selected so as to
reduce the errors in HR2, allowing us to discriminate between the two
favored spectral models: thermal or non-thermal (power-law). The net
counts in the three bands for each MSP are listed in Table 1, while
the resulting X-ray color-color diagram is shown in Figure 2.  For
comparison, we have also plotted values of HR1 and HR2 for simple
absorbed models of blackbody, neutron star hydrogen atmosphere
\citep*{Rom87,Zavlin96,Lloyd03}, and power-law spectra for a range of
temperatures and photon indices, respectively.

The \textit{Chandra} ACIS-I observations of 47 Tuc revealed that the
X-ray colors of the majority of the MSPs are consistent with those of
soft, presumably thermal, sources with blackbody temperatures of
$kT\sim0.22$ keV and emission radii of $R < 0.6$ km \citep{Grind02}.
The X-ray color-color diagram for the ACIS-S data (Fig. 2) shows that,
with the exception of 47 Tuc J and W, the MSPs are indeed soft sources
although the positions of the MSPs on the diagram are suggestive of a
more complex spectrum, such as a composite thermal and non-thermal
(power-law) spectrum or a multi-temperature thermal spectrum. As
\citet{Grind02} and \citet{Ran04} argue, although consistent with the
X-ray colors of the MSPs, a pure thermal bremsstrahlung spectrum is
ruled out as it would require unrealistically large plasma densities,
which would be inconsistent with the observed dispersion measures.

The fivefold improvement in counting statistics, resulting from the
longer exposure time and the better soft response of the ACIS-S
compared to the ACIS-I observations, permits a more precise
determination of the nature of the X-ray emission via spectral fitting
for the 47 Tuc MSPs. For this purpose, we used the total counts
extracted for each MSP from the four observations to generate source
and background spectra, along with the corresponding response matrix
and auxiliary files.  The MSP pairs 47 Tuc F and S, and G and I, were
treated as single sources as the individual spectra could not be
separated.  For the fainter MSPs, the extracted counts were grouped
into energy bins containing at least 10 counts, while for the brighter
MSPs each bin contained at least 15 counts. The subsequent
least-squares fits to the spectra were performed using the
XSPEC\footnote{http://heasarc.gsfc.nasa.gov/docs/xanadu/xspec/}
software package and were restricted to photon energies between 0.3
and 8 keV. In the fitting process, we considered the following
spectral models as plausible physical descriptions of the MSP
emission: a pure blackbody (BB), an unmagnetized neutron star hydrogen
atmosphere \citep[NSA;][]{Lloyd03}, a pure power-law (PL), a composite
thermal and power-law (BB+PL or NSA+PL), and a two-temperature thermal
model (BB+BB or NSA+NSA). Throughout this analysis the hydrogen column
density was fixed at the nominal value for 47 Tuc, $N_{\rm
H}=(1.3\pm0.3)\times10^{20}$ cm$^{-2}$
\citep{Gratt03,Pred95,Card89}. For the NSA model we assumed a value of
$z_{g} =
\big[1-\left(2GM_{NS}\right)/\left(c^2R_{NS}\right)\big]^{-1/2}-1=0.31$
for the gravitational redshift at the NS surface, appropriate for a
$M_{\rm NS}=1.4$ M$_{\odot}$, $R_{\rm NS}=10$ km star.  Thus, the only
free parameters for the BB and NSA fits were the effective temperature
and BB/NSA normalization, while for the PL model the free parameters
were the photon index and the normalization.

\begin{deluxetable*}{lcccccl}[]
\tabletypesize{\small} 
\tablecolumns{7} 
\tablewidth{0pc}
\tablecaption{Best fit spectral models and unabsorbed fluxes for the
47 Tuc MSPs.}

\tablehead{ \colhead{} & \colhead{Spectral} &
\colhead{$T_{\rm eff}$} & \colhead{$R_{\rm eff}$\tablenotemark{b}} & \colhead{Photon} & \colhead{$\chi^2_{\nu}$/dof} & \colhead{$F_{\rm X}$ (0.3-8 keV)} \\
\colhead{MSP} & \colhead{Model\tablenotemark{a}} & \colhead{($10^6$ K)} & \colhead{(km)} &
\colhead{Index} & & \colhead{($10^{-15}$ ergs cm$^{-2}$ s$^{-1}$)} }

\startdata
 C & BB & $2.02\pm0.18$ & $0.11\pm0.08$ & - & $0.29/3$ & $0.71^{+0.14}_{-0.26}$ \\
 D & BB & $2.20\pm0.17$ & $0.13\pm0.06$  & -  & $1.96/8$ & $1.37^{+0.18}_{-0.35}$ \\
 E & BB & $1.75\pm0.09$ & $0.28\pm0.17$ & - & $1.60/6$ & $2.08^{+0.26}_{-0.36}$ \\
 F+S & BB & $2.19\pm0.09$ & $0.22\pm0.11$ & - & $1.05/10$ & $3.80^{+0.31}_{-0.56}$  \\
 G+I & BB & $2.21\pm0.12$ & $0.18\pm0.10$ & - & $0.96/12$ & $2.54^{0.19+}_{-0.48}$ \\
 H & BB & $1.93\pm0.13$ & $0.17\pm0.11$ & -  & $0.55/6$ & $1.30^{+0.12}_{-0.33}$ \\
 J & BB+PL & $1.73\pm0.21$ & $0.22\pm0.17$ & $1.00\pm0.56$ & $1.07/5$ & $4.77^{+0.95}_{-1.51}$ ($3.28^{+0.19}_{-2.32}$) \\
 L\tablenotemark{c} & BB & $2.27\pm0.10$ & $0.20\pm0.10$ & - & $2.04/11$ & $3.54^{+0.32}_{-0.41}$ \\
 M & BB & $2.22\pm0.18$ & $0.11\pm0.07$  & -  & $0.40/4$ & $1.01^{+0.14}_{-0.30}$ \\
 N & BB & $2.07\pm0.18$ & $0.13\pm0.09$ & - & $1.94/5$ & $0.98^{+0.17}_{-0.30}$ \\    
 O\tablenotemark{c} & BB+PL & $1.76\pm0.15$ & $0.28\pm0.18$ & $1.33\pm0.79$  & $1.21/10$ & $4.44^{+1.44}_{-0.86}$ ($2.23^{+0.24}_{-1.73}$) \\
 Q & BB & $2.24\pm0.20$ & $0.11\pm0.07$ & - & $1.17/5$ & $1.00^{+0.13}_{-0.30}$ \\
 R & BB & $2.51\pm0.16$  & $0.15\pm0.08$ & - & $2.54/7$ & $2.87^{+0.17}_{-0.56}$     \\
 T & BB & $1.56\pm0.16$ & $0.19\pm0.17$ & - & $2.09/2$ & $0.63^{+0.13}_{-0.26}$ \\
 U & BB & $2.76\pm0.22$ & $0.08\pm0.06$ & - & $0.55/6$ & $1.32^{+0.12}_{-0.33}$ \\
 W & BB+PL & $1.52\pm0.28$ & $0.29\pm0.29$  & $1.36\pm0.24$ & $1.22/14$ & $10.90^{+0.40}_{-2.61}$ ($9.55^{+1.75}_{-1.44}$) \\
 Y & BB & $2.24\pm0.18$ & $0.11\pm0.07$ & - & $1.16/4$ & $1.03^{+0.10}_{-0.28}$ \\
\enddata

\tablenotetext{a}{PL is a power-law and BB is a single temperature
blackbody model. For all MSPs the hydrogen column density was fixed
at $N_{\rm H}=(1.3\pm0.3)\times10^{20}$ cm$^{-2}$. In the last column,
the values in parentheses for J, O, and W represent the flux in the PL
component. All uncertainties quoted are $1\sigma$.}

%\tablenotetext{b}{$R_{\rm eff}$ calculated assuming a distance of 4.5 kpc.}

%\tablenotetext{c}{Spectrum may be contaminated by background/neighboring source(s).}

\end{deluxetable*}

\begin{deluxetable*}{lcccccl}
\tabletypesize{\small} 
\tablecolumns{7} 
\tablewidth{0pc}
\tablecaption{Best fit spectral models and unabsorbed fluxes for the
47 Tuc MSPs.}

\tablehead{ \colhead{} &
\colhead{Spectral} & \colhead{$T_{\rm eff}$} & \colhead{$R_{\rm eff}$\tablenotemark{b}}
& \colhead{Photon} & \colhead{$\chi^2_{\nu}$/dof} &
\colhead{$F_{\rm X}$ (0.3-8 keV)} \\ \colhead{MSP} & \colhead{Model\tablenotemark{a}} & \colhead{($10^6$ K)} & \colhead{(km)} & \colhead{Index} & & \colhead{($10^{-15}$ ergs cm$^{-2}$ s$^{-1})$}}

\startdata
 C & NSA & $1.12^{+0.23}_{-0.19}$  & $0.57\pm0.56$ & - & $0.68/3$ & $0.73^{+0.09}_{-0.29}$ \\
 D & NSA & $1.29^{+0.20}_{-0.17}$ & $0.61\pm0.51$ & - & $1.40/8$ & $1.54^{+0.05}_{-0.62}$ \\
 E & NSA & $0.88^{+0.12}_{-0.11}$ & $1.75\pm1.41$ & - & $1.03/6$ & $2.28^{+0.05}_{-0.75}$ \\
F+S & NSA & $1.27^{+0.10}_{-0.09}$ & $1.03\pm0.64$ & - & $0.77/10$ & $4.09^{+0.14}_{-0.62}$  \\
G+I & NSA & $1.28^{+0.12}_{-0.12}$ & $0.84\pm0.58$ & - & $0.60/12$ & $2.78^{+0.16}_{-0.50}$ \\
 H & NSA & $1.04^{+0.16}_{-0.14}$ & $0.94\pm0.79$ & - & $0.59/6$ & $1.39^{+0.03}_{-0.68}$ \\
 J & NSA+PL & $0.89^{+0.20}_{-0.16}$ & $1.43\pm1.40$ & $1.0$ & $1.00/6$ & $4.69^{+1.07}_{-1.56}$ ($3.09^{+1.99}_{-0.38}$) \\
 L\tablenotemark{c} & NSA & $1.42^{+0.12}_{-0.12}$ & $0.78\pm0.50$ & - & $2.09/11$ & $3.70^{+0.24}_{-0.72}$\\
 M & NSA & $1.27^{+0.19}_{-0.17}$ & $0.53\pm0.45$  & - & $0.45/4$ & $1.08^{+0.08}_{-0.40}$ \\
 N & NSA & $1.20^{+0.22}_{-0.19}$ & $0.61\pm0.57$ & - & $1.38/5$ & $1.09^{+0.03}_{-0.50}$ \\
 O\tablenotemark{c} & NSA+PL & $0.98^{+0.17}_{-0.15}$ & $1.48\pm1.27$ & $1.3$  & $1.34/11$ & $4.55^{+1.16}_{-1.49}$ ($1.91^{+2.62}_{-0.17}$) \\
 Q & NSA & $1.30^{+0.26}_{-0.21}$ & $0.51\pm0.46$ & - & $0.78/5$ & $1.10^{+0.03}_{-0.43}$ \\
 R & NSA & $1.54^{+0.19}_{-0.17}$ & $0.60\pm0.45$ & - & $1.60/7$ & $3.19^{+0.13}_{0.92}$ \\
 T & NSA & $0.80^{+0.19}_{-0.15}$  & $0-2.4$ & - & $1.52/2$ & $0.67^{+0.03}_{-0.32}$ \\
 U & NSA & $1.82^{+0.27}_{-0.24}$ & $0.28\pm0.22$ & - & $0.49/6$ & $1.40^{+0.07}_{-0.60}$ \\
 W & NSA+PL & $0.94^{+0.34}_{-0.25}$ & $0-2.1$  & $1.15\pm0.34$ & $1.17/14$ & $11.32^{+2.45}_{-2.35}$ ($9.21^{+2.11}_{-1.99}$) \\
 Y & NSA & $1.35^{+0.20}_{-0.17}$ & $0.47\pm0.40$ & - & $0.98/4$ & $1.10^{+0.07}_{0.33}$ \\
\enddata

\tablenotetext{a}{PL is a power-law and NSA is a single temperature
NS hydrogen atmosphere model. For 47 Tuc J and O, the PL index has been fixed at the nominal value obtained for the BB+PL fit as the fit parameters cannot be reliably constrained for the NSA+PL with the PL index as a free parameter.}

\tablenotetext{b}{$R_{\rm eff}$ calculated assuming a distance of 4.5 kpc.}

\tablenotetext{c}{Spectrum may be contaminated by background/neighboring source(s).}

\end{deluxetable*}

From the spectral fits we find that many of the 47 Tuc MSP spectra are
well described by a purely thermal spectrum (BB or NSA).  For the BB
model the fits yielded effective temperatures and radii of $T_{\rm
eff}\sim(1.5-3.0)\times10^6$ K and $R_{\rm eff}\sim0.1-0.5$ km,
respectively, while in the case of the NSA model we obtained $T_{\rm
eff}\sim(0.8-1.8)\times10^6$ K and $R_{\rm eff}\sim0.3-3$ km. Tables 2
and 3 list the best fit model, the corresponding values for the fitted
parameters, and the unabsorbed flux for each MSP for the BB and NSA
model fits, respectively.  Figure 3 shows representative thermal
spectra for the 47 Tuc MSPs.

It is apparent from the values of $\chi^2_{\nu}$ listed in Tables 2
and 3, as well as Figure 3, that the spectra of several MSPs are
poorly fit by a single temperature model, especially at energies above
$\sim$1.5 keV. In addition, we find the fits to be unacceptable for a
PL or a composite PL+BB/PL+NSA model as well. Motivated by the results
for the nearby MSPs J0437--4715 \citep{Zavlin02} and J0030+0451
\citep{Beck02} we attempted fitting a two-temperature polar cap model
to the brighter thermal-emitting 47 Tuc MSPs, namely, D, E, L, and
R. We found that a two-temperature spectrum can account for the excess
at $>$1.5 keV apparent in Figure 3, resulting in an improved fit,
although the limited count statistics do not allow us to reliably
constrain the emission areas. Nonetheless, if we fix the two
temperatures to values comparable to those of J0437--4715 and
J0030+0451 we obtain very similar results for the effective radii of
the 47 Tuc MSPs to those obtained for the two field
MSPs. Specifically, in the case of the NSA model, for $T_1= 2.1 \times
10^6$ K and $T_2= 0.54 \times 10^6$ K \citep{Zavlin02} the fits
yielded $R_1\sim 0.3$ km and $R_2\sim 2$ km, respectively, while for a
BB model with $T_1= 3.2 \times 10^6$ K and $T_2= 1.4 \times 10^6$ K
\citep{Beck02}, we obtain $R_1\sim50$ m and $R_2\sim300$ m.  If we
extend the analogy further, it seems plausible that the thermal
emission from all 47 Tuc MSPs consists of two components as well.
This imples that the derived values of $T_{\rm eff}$ and $R_{\rm eff}$
listed in Tables 2 and 3 for the single component fit represent
averages, weighted based on the relative temperatures and areas of the
two emitting regions.

\begin{figure*}
\begin{center}
\includegraphics[width=0.89\textwidth]{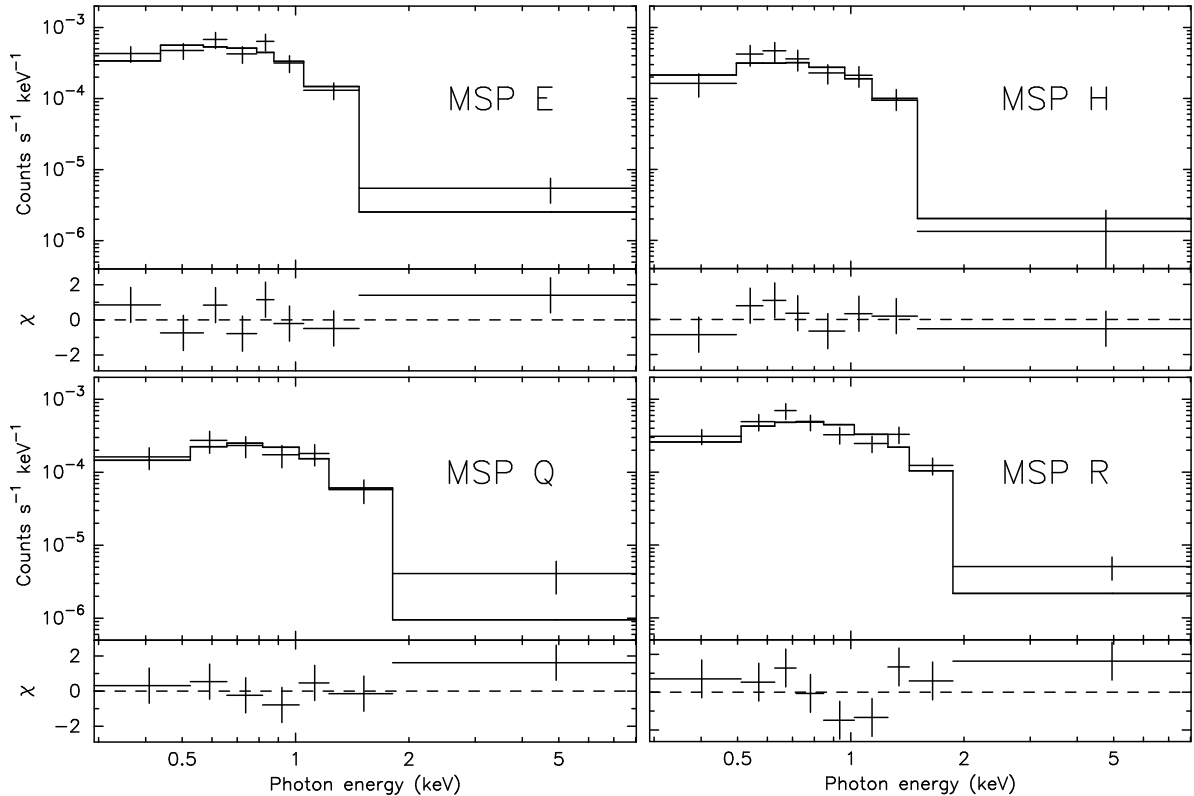}
\caption{Representative X-ray spectra and best fit single-temperature thermal models (\textit{solid line}) for the 47 Tuc MSPs. The lower panel for each MSP shows the best fit residuals. See the electronic edition of \textit{The Astrophysical Journal} for the spectra of all 47 Tuc MSPs. }
\end{center}
\end{figure*}

For the fainter MSPs, such as 47 Tuc C, M, N, Q, and T, a pure
power-law model, with a fairly steep spectrum ($\Gamma\gtrsim 2.5$),
also yields an acceptable fit. However, we believe that this likely
stems from the fact that a two-temperature spectrum appears
power-law-like in the energy range of interest \citep[see e.g. Fig. 2
in][]{Zavlin02}.  We also note that a broken power-law model, with a
break at $E_{b}\approx1.0$ keV, is also a good description of the
observed emission from the 47 Tuc MSPs. In principle, such a spectrum
could arise due to a deficit of radiating high-energy particles in the
pulsar magnetosphere. A break in the spectrum is observed in young
normal pulsars though at much higher energies \citep[a few GeV, see
e.g.,][]{Tho96}.  However, evidence against a purely non-thermal model
comes from optical observations of PSR J0030+0451, an MSP with a
qualitatively similar spectrum to those of the 47 Tuc pulsars, which
show that such a model grossly overestimates (by a factor of
$\sim$500) the optical flux upper limits when extrapolated to lower
energies \citep{Kop03}.

For the most luminous MSPs in the sample, namely 47 Tuc J, O, and W, a
composite NSA+PL or BB+PL model yields an acceptable fit, with the PL
component contributing $\sim$70\%, $\sim$50\%, and $\sim$75\% of the
total flux, respectively.  The spectral fits, shown in Figure 4, were
found to be acceptable for photon indices $\Gamma \sim 1.0-1.4$. We
note that although an F-test does not indicate that a composite
spectrum is statistically prefered over a pure PL (but with a steeper
photon index of $\Gamma\sim1.5-2.0$), the fainter thermal component is
very likely genuine, given that the inferred values for $T_{\rm eff}$
and $R_{\rm eff}$ are very similar to those of the other 47 Tuc MSPs.
The non-thermal X-rays from MSP W, and probably MSP J and O as well,
most likely originate in an intrabinary shock (see \S4) although for J
and O at least a portion of this emission may originate in the pulsar
magnetosphere.  For 47 Tuc O, which is near the crowded center of the
globular cluster, the PL component may also be due to a background
X-ray source. The spectra of all other 47 Tuc MSPs may, in fact,
contain a faint magnetospheric emission component, which is
undetectable in our observations due to the limited number of counts.

It is important to note that, in order to model the NSA and BB spectra
of MSPs properly, one needs to take into account the rotation of the
NS, which induces a modulation in the projected area of the polar
cap(s) at the spin period and, in the case of a NSA, in the observed
spectrum as well due to the energy dependence of the limb darkening
effects of a NS atmosphere \citep[see
e.g.][]{Rom87,Zavlin96,Lloyd03}. Therefore, the derived temperatures
and radii represent time-averaged values integrated over the rotation
period of the pulsar. As a consequence, the spectral fits may
underestimate the true flux by up to a factor of $\sim$2, even if we
consider the effects of gravitational redshift and bending of light
near the NS surface \citep*[see
e.g.][]{MTW70,Pech83,Belo02}. Unfortunately, in order to account for
these effects, one requires prior knowledge of the relative
orientation of the pulsar spin and magnetic axes and the line of
sight, which is unavailable.

\begin{figure}
\begin{center}
\includegraphics[width=0.47\textwidth]{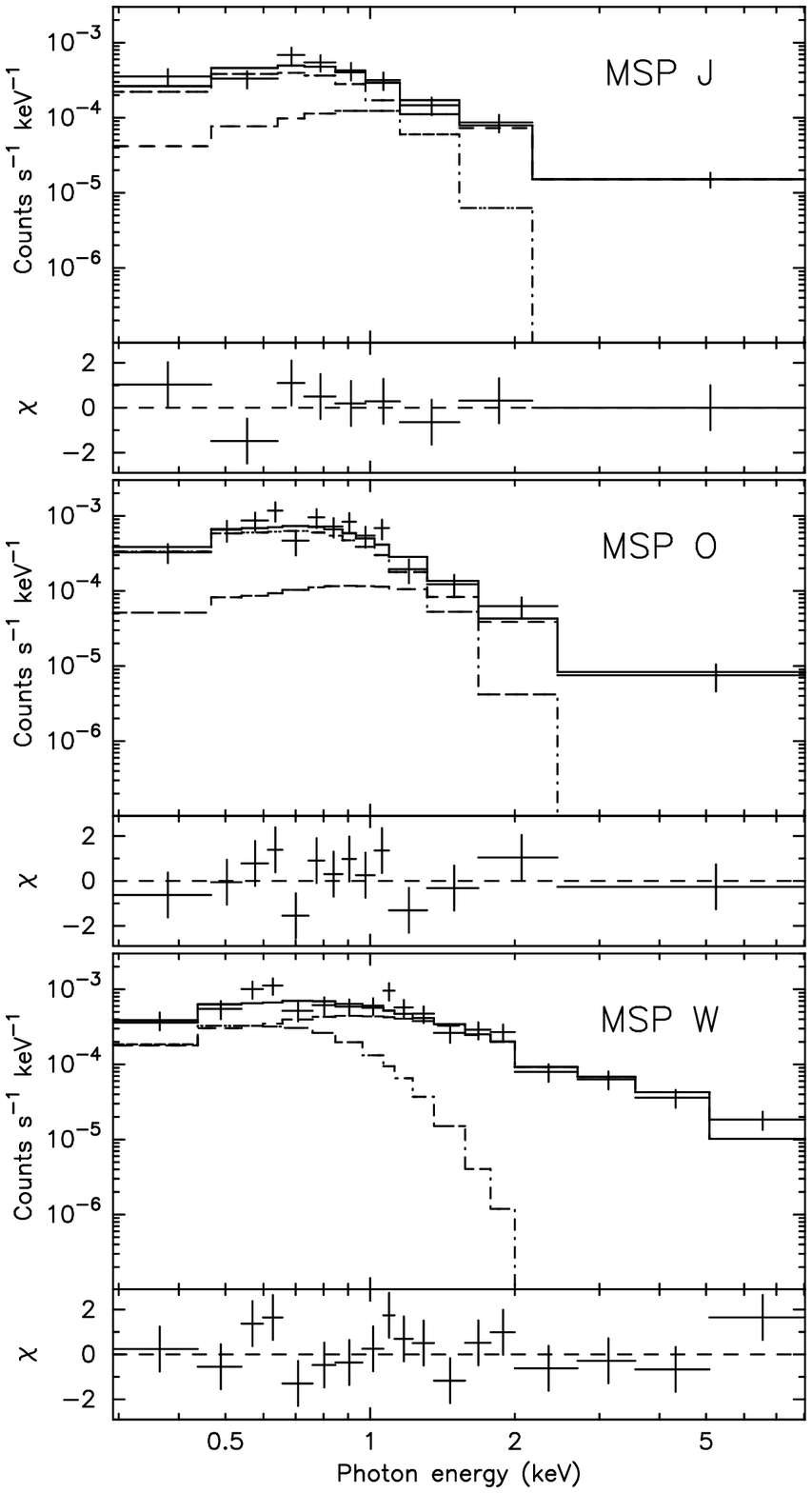}
\end{center}
\caption{The X-ray spectra of 47 Tuc J, O, and W, which exhibit
non-thermal emission. The solid line shows the best-fit model
consisting of a power-law (dashed line) and a thermal (dot-dashed
line) component. The lower panel for each MSP shows the best fit
residuals.}
\end{figure}

\section{VARIABILITY ANALYSIS}

Although the limited time resolution of the ACIS-S observations (3.2
seconds) precludes a timing analysis at the millisecond level, these
observations in conjunction with the 2000 ACIS-I observations, permit
us to investigate the temporal behavior of the MSP emission over a
large range of timescales from hours to years.

As expected, for the bulk of the sample we find no statistically
significant X-ray variability. For MSP O, however, there seems to be a
gradual decrease in the photon count rate, by a factor of $\sim$$2$ in
total, especially at medium energies (0.6--1.5 keV), from the second
to the fourth ACIS-S observation. There is no known physical mechanism
which could account for such a substantial variation of the flux on
timescales of days from an old MSP such as 47 Tuc O. On the other
hand, such behavior is typical of cataclysmic variables and
chromospherically active binaries with $L_{X}\sim10^{30-31}$ ergs
s$^{-1}$. Thus, we conclude that this variation is most likely due to
blending of MSP O with such an X-ray source. This seems likely as O is
at the heart of the cluster (see Fig. 1), where the number density of
X-ray sources is large.  The MSP R is found to be marginally variable
between observations, while MSP U exhibits possible variability within
the third observation \citep{Heinke05}. In both cases, the apparent
variability is likely spurious considering the probability of spurious
detection and the number (19) of trials.

For the 12 binary MSPs, we can also examine whether there exist any
variations in the X-ray flux as a function of orbital position. In
this sense, 47 Tuc J, O, R, and W are of particular interest as they
are in very compact binaries with periods of 2.9, 3.3, 1.6, and 3.2
hours, respectively. The former three MSPs are bound to very low mass
degenerate companions ($m_c \sim 0.03$ M$_{\odot}$) and are
eclipsed regularly at radio wavelengths for 10-25\% of their orbits
\citep{Camilo00,Freire05}.  MSP W, on the other hand, has a
significantly more massive ($m_c \gtrsim 0.13$ M$_{\odot}$)
main-sequence secondary \citep{Edm02} and undergoes eclipses for about
35--40\% of the entire orbit \citep{Freire05}.  For this analysis, the
observations for the binary MSPs were first folded at the binary
period using the latest values for the orbital periods $P_{b}$ and the
epochs of ascending node $T_0$ \citep[][Freire et al. unpublished
data]{Freire03}.  The net counts were then grouped in phase bins sized
so as to allow detection of any large-amplitude variations in the
X-ray photon count rate at the eclipse phases.  For the same purpose,
we also performed Kolmogorov-Smirnov and Cramer-von Mises tests on the
folded but unbinned data, using an integer number of binary orbits.

For 47 Tuc J, O, and R we find no detectable variation in the X-ray
count rate at any orbital phase and are only able to set very crude
upper limits on the amplitude of variation of $\lesssim$60-80\%.  In
contrast, 47 Tuc W exhibits dramatic variations in the X-ray flux
(significant at the 99.9\% level) as a function of orbital phase. This
behavior can be naturally explained by the existence of a swept-back
shocked stream of gas formed by interaction of the pulsar wind with
matter from the irradiated companion issuing through the inner
Lagrange point.  The properties of this intriguing system and its
possible connection to low mass X-ray binaries are described in depth
by \citet*{Bog05}.

\begin{deluxetable}{lcc}

\tabletypesize{\small} 
\tablecolumns{3} 
\tablewidth{0pc}
\tablecaption{Derived $\dot{E}$ and $L_X$ for the 47 Tuc MSPs.}

\tablehead{ \colhead{ } & \colhead{$\dot{E}$\tablenotemark{a}} & \colhead{$L_X$\tablenotemark{b}} \\
\colhead{MSP} & \colhead{($\times 10^{34}$ ergs s$^{-1}$)} & \colhead{($\times 10^{30}$ ergs s$^{-1}$)}}

\startdata

C  &  $0.05_{-0.05}^{+0.12}$ & 	$2.6_{-2.4}^{+2.9}$   \\          
D  &  $0.67_{-0.21}^{+0.27}$ & 	$4.8_{-2.7}^{+2.4}$   \\    
E  &  $3.12_{-0.79}^{+0.79}$ & 	$8.9_{-7.3}^{+9.8}$   \\  
F\tablenotemark{c}  &  $4.09_{-3.16}^{+3.16}$ & $5.5_{-4.2}^{+6.8}$  \\ 
G(+I)  &  $<$1.72 	     & 	$9.3_{-7.3}^{+9.3}$   \\
H\tablenotemark{d}  &   -    & 	$4.8_{-4.0}^{+5.4}$   \\ 
I  &  $<$7.14                & 	-	              \\ 
J  &  $3.22_{-1.61}^{+1.61}$ &	$5.6_{-5.0}^{+4.6}$   \\  
L  &  $1.04_{-1.04}^{+1.77}$ &	$13.8_{-9.8}^{+12.1}$   \\  
M  &   -                     &	$3.8_{-3.3}^{+5.2}$   \\ 
N  &  $1.87_{-1.06}^{+1.49}$ & 	$3.9_{-3.4}^{+5.5}$   \\ 
O  &  $3.12_{-1.13}^{+0.85}$ &  $9.3_{-7.6}^{+8.1}$   \\   
Q  &  $1.82_{-0.12}^{+0.12}$ & 	$3.9_{-3.4}^{+5.5}$   \\  
R  &  $2.84_{-2.28}^{+3.10}$ & 	$11.0_{-8.6}^{+12.1}$   \\   
S\tablenotemark{c}  &  $2.27_{-2.27}^{+0.49}$ & $8.9_{-6.5}^{+9.9}$   \\   
T  &  $1.09_{-0.69}^{+0.41}$ & 	$2.9_{-2.7}^{+3.1}$   \\   
U  &  $3.98_{-0.21}^{+0.21}$ &  $5.1_{-4.4}^{+7.8}$   \\   
W\tablenotemark{d}  &   -    &  $5.9_{-5.9}^{+3.3}$    \\
Y  &  $4.82_{-4.49}^{+6.12}$ &  $3.9_{-3.4}^{+5.5}$    \\

\enddata

\tablenotetext{a}{Values for $\dot{E}$ derived assuming a King model for 47 Tuc using the following parameters: distance D=4.5 kpc, central velocity dispersion $v_z(0)=11.6$ km s$^{-1}$, core radius $r_c=23.1''$, and best fit constant gas density $n_e=0.1$ cm$^{-3}$.}

\tablenotetext{b}{Values calculated using $L_X=4\pi R^2\sigma_{SB}T^4$, with $R$ and $T$ taken from Table 2, and corrected for gravitational redshift assuming $1+z_g=1.31$. For MSP G, $L_X$ is for G and I combined.}

\tablenotetext{c}{$L_X$ apportioned based on procedure described in \S2.}

\tablenotetext{d}{No reliable measurements of $\dot{P}$.}

\end{deluxetable}

\begin{figure}[!t]
\begin{center}
\includegraphics[width=0.42\textwidth]{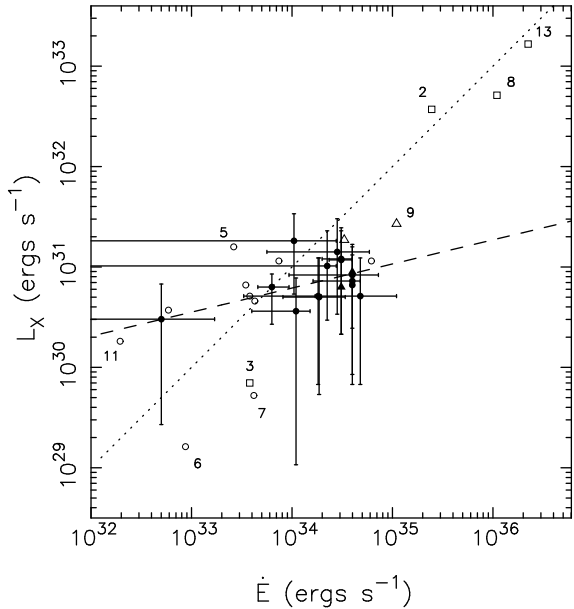}
\caption{X-ray luminosity versus spin-down luminosity ($\dot{E}=4\pi^2
I \dot{P}_{i}/P^3$) for the MSPs in 47 Tuc (filled symbols) and all
other MSPs detected in X-rays (open symbols).  The X-ray luminosities
due to polar cap (thermal), magnetospheric, and shock emission are
shown as circles, squares, and triangles, respectively. For 47 Tuc J
and O, and J0437-4715 the individual emission components are plotted
separately. The dashed line corresponds to the best fit for the 47 Tuc
MSPs with thermal spectra, while the dotted line shows the linear
relation $L_{\rm X}\propto 10^{-3}\dot{E}$. The numerical labels for
the open symbols correspond to the values listed in the first column
of Table 5.}
\end{center}
\end{figure}

\section{X-RAY LUMINOSITY VERSUS PULSAR PARAMETERS}

\subsection{X-ray versus Spindown Luminosity}

Rotation-powered pulsars, including MSPs, appear to exhibit a linear
relation between their X-ray luminosity $L_X$ and rotational spindown
luminosity $\dot{E}=4\pi^2 I \dot{P}_{i}/P^3$, where $I$ is the NS
moment of inertia typically assumed to be $10^{45}$ g cm$^2$, with
$L_{X}\sim 10^{-3} \dot{E}$ \citep{Beck97,Poss02}. However,
\citet{Grind02} have shown that the dependence of $L_{X}$ on $\dot{E}$
for the 47 Tuc MSPs may be significantly flatter, with
$L_{X}\propto\dot{E}^{0.5}$.

Using the present data, we can re-examine the $L_{X}-\dot{E}$ relation
for the 47 Tuc MSPs. However, the measured values of $\dot{P}$ and,
hence, $\dot{E}$ for these MSPs are significantly affected by
acceleration in the gravitational potential of the globular
cluster. In order to obtain the intrinsic period derivative
$\dot{P}_i$, we first determined the three-dimensional position of each
MSP relative to the cluster center by considering the projected radius
and assuming that the observed differences in dispersion measure
within the MSP sample are solely due to a spread in distance along the
line of sight for an assumed uniform intra-cluster plasma
\citep{Freire01b}.  The cluster acceleration term for each MSP was
then computed assuming a King model for the cluster using a central
velocity dispersion of $v_z(0)=11.6$ km s$^{-1}$, core radius
$r_c=23.1''$, and distance $D=4.5$ kpc
\citep{Mey86,Freire01b}. Subtracting the acceleration term from the
observed $\dot{P}$ yields the intrinsic spindown rate $\dot{P}_i$,
which was then used to compute $\dot{E}$. The resulting values and
their uncertainties are listed in Table 4. In this calculation we have
not included MSP H, which exhibits anomalous variations in $\dot{P}$,
and MSP W, which currently has no measured $\dot{P}$.

Figure 5 shows the plot of $L_{X}$ versus $\dot{E}$ for the sample of
MSPs in 47 Tuc based on the values listed in Table 4.  The distance
used to compute $L_{X}$ for the 47 Tuc MSPs was taken to be 4.5
kpc. For the MSP pair F and S, the combined $L_X$ was apportioned based
on the procedure described in \S 2. In the case of G and I the
individual contribution of each MSP to the observed $L_X$ is unknown.
Also plotted are all MSPs detected in X-rays in the field of the
Galaxy as well as in the globular clusters M4, M28, M30, NGC 6397, and
NGC 6752. The latest values for the parameters of these MSPs are
summarized in Table 5. Here, we have chosen to compute the X-ray
luminosity $L_{X}=4 \pi D^2 F_{X}$ in a broad energy band (0.1--10
keV) instead of the \textit{ROSAT} band (0.1--2.4 keV) in order to
take into account the substantial portion of the total flux (up to
$\sim$80\%) that is present beyond 2.4 keV for the MSPs with hard
spectra.

In this analysis, we have classified the MSPs based on the three
distinct types of X-ray emission: (i) non-thermal magnetospheric, (ii)
non-thermal shock, and (iii) thermal polar cap emission. In the latter
category we have also included the MSPs with undetermined spectral
properties as they have very similar luminosities to the pulsars with
thermal spectra.  In addition, all thermal luminosities have been
corrected for the gravitational redshift, assuming $z_{g} = 0.31$. For
the MSPs with known multi-component spectra, such as PSR J0437--4715
and 47 Tuc J and O, the emission has been decomposed into the
individual constituents.

The thermal luminosities of the 47 Tuc MSPs alone follow the relation
$\log L_{X}=(0.24\pm1.10)\log \dot{E}+(24\pm37)$. In this fit we have
taken into account the errors in both $L_X$ and $\dot{E}$ listed in
Table 4.  As evident in Figure 5, due to the large uncertainties, the
47 Tuc sample appears consistent with both the linear relation and a
much flatter trend, such as $L_{X}\propto\dot{E}^{0.5}$, predicted by
the \citet{Hard02a} polar cap heating model. It is interesting to note
that both field and cluster MSPs with thermal spectra occupy the same
region of the $L_{X}-\dot{E}$ diagram. This suggests that in terms of
X-ray properties the two populations are indistinguishable.

Curiously, despite the fundamentally different emission mechanisms,
all MSPs appear to follow the linear relation $L_X\propto
10^{-3}\dot{E}$, and have an efficiency of converting $\dot{E}$ to
$L_{X}$ covering a surprisingly narrow range, with $L_{X}/\dot{E}\sim
10^{-4}-10^{-3}$.  This is not expected given that in systems such as
47 Tuc J, O, and W, B1957+20, and J1740--5340, whose emission most
likely originates in an intrabinary shock, $L_{X}$ is determined, in
part, by parameters unrelated to the physics of the pulsar, such as
the binary separation, the properties of the companion star, and the
volume of the shocked region.  The small range in $L_{X}/\dot{E}$ may
stem from the fact that in all cases the high-energy emission is,
ultimately, driven by the same underlying particle acceleration and
pair production processes occurring in the pulsar
magnetosphere. However, the fact that MSPs with
$\dot{E}\lesssim10^{35}$ ergs s$^{-1}$ appear to have thermal spectra,
whereas those with $\dot{E}\sim10^{36}$ ergs s$^{-1}$ show strong
non-thermal emission, suggests a profound difference in the conditions
in the pulsar magnetosphere (see \S6.3).

\subsection{X-ray Luminosity versus Magnetic Field Strength}

Next, we consider how $L_{X}$ depends on the pulsar magnetic field as
it governs the physical processes responsible both for magnetospheric
emission, polar cap heating, and, to some degree, shock emission. In
Figure $6a$, we plot $L_{X}$ versus $B_{\rm
lc}=9.35\times10^{5}(P/10^{-3}~{\rm s})^{-5/2}(\dot{P}/10^{-20})^{1/2}$ G, the magnetic field at
the light cylinder ($r_{c}=cP/2\pi$) for a simple dipole.  The MSPs
with thermal spectra are consistent with $\log L_{X}=(0.39\pm0.54)\log
B_{\rm lc}+(28\pm3)$. As with the $L_X-\dot{E}$ relation, the fit is
not particularly constraining due to the large uncertainties.
Nonetheless, as noted by \citet{Saito97}, it is striking that
B1821--24 and B1937+21, which exhibit very luminous non-thermal
(magnetospheric) X-ray emission, have the highest magnetic fields at
the light cylinder, with $B_{\rm lc}>10^{5.5}$ G.  It may be the case
that these MSPs do, in fact, have heated polar caps but the resulting
emission is several orders of magnitude fainter than the
magnetospheric emission.  Alternatively, the energetics of the system
may not favor substantial polar cap heating \citep[see,
e.g.][]{Hard02a}.  Pulsars with $B_{\rm lc}<10^5$ G, on the other
hand, appear to have X-ray emission that is typically either thermal
or of indeterminate (but probably thermal) character.

In the case of the dipole magnetic field at the surface $B_{\rm
surf}=3.2\times10^{19}(P\dot{P})^{1/2}$ G, the $L_{X}-B_{\rm surf}$
relation (Fig. $6b$) seems even less defined with a large scatter for
both the complete MSP sample and the 47 Tuc MSPs. The thermal $L_X$
for the 47 Tuc MSPs is best fit by $\log L_{X}=(0.36\pm1.27)B_{\rm
surf}+(22\pm37)$.  The apparent lack of correlation between $L_{X}$
and $B_{\rm surf}$ may either indicate that the small scale magnetic
field structure near the surface of the polar caps deviates from that
of a pure dipole \citep[see e.g.,][]{Zhang03} or that both the thermal
and non-thermal X-ray emission properties of MSPs are determined by
the conditions much higher in the magnetosphere (e.g., at the light
cylinder).

\begin{figure}[t!]
\begin{center}
\includegraphics[width=0.46\textwidth]{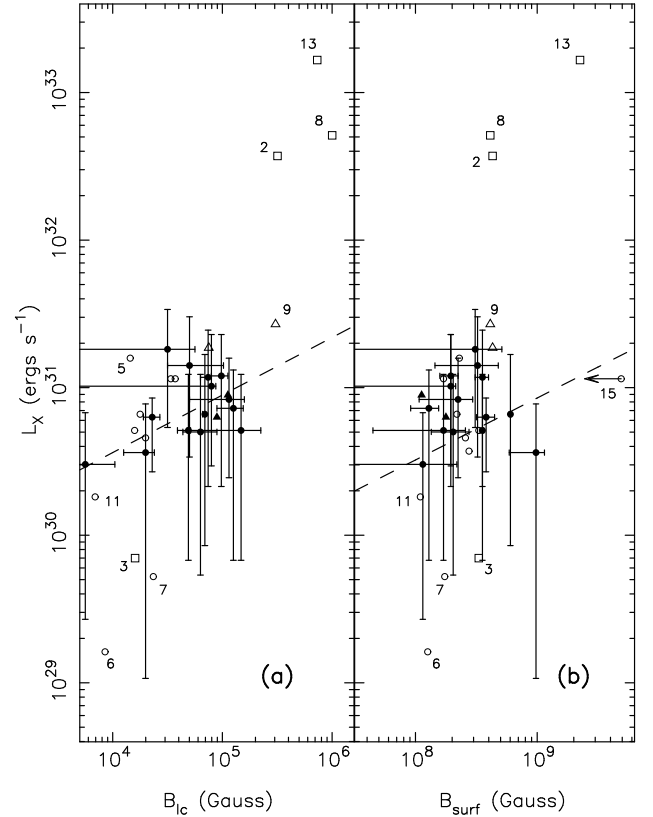}
\caption{(\textit{a}) $L_{\rm X}$ versus magnetic field strength at
the light cylinder ($B_{\rm lc}$) for the MSPs in 47 Tuc (filled
symbols) and all other MSPs detected in X-rays (open
symbols). (\textit{b}) $L_{\rm X}$ versus surface magnetic field strength
($B_{\rm surf}$). The dashed line corresponds to the best fit for the
47 Tuc MSPs with thermal spectra. The various symbols are defined as
in Figure 5.}
\end{center}
\end{figure}

\begin{deluxetable*}{llccccc}
\tabletypesize{\small}
\tablecolumns{7} 
\tablewidth{0pc}
\tablecaption{Parameters for MSPs detected in X-rays outside of 47 Tuc.}
\tablehead{
\colhead{ } & \colhead{ }  & \colhead{$P$} & \colhead{$\dot{P}_{\rm i}$} & \colhead{$D$} & \colhead{$F_{\rm X}$ (0.1-10 keV)\tablenotemark{b}} & \\ \colhead{ } &
\colhead{MSP\tablenotemark{a}} & \colhead{(ms)} & \colhead{($10^{-20}$)} & \colhead{(kpc)} & \colhead{($10^{-13}$ ergs cm$^{-2}$ s$^{-1}$)} & \colhead{References}}

\startdata 
1 & J0030+0451 & 4.86 & $<$1.0\tablenotemark{c} & 0.3 & 3.6 & 1, 2 \\
2 &J0218+4232 & 2.32 & 7.78 & 2.7 & 4.2 & 3, 4 \\ 
3 & J0437--4715 & 5.76 & 1.86 & 0.139\tablenotemark{d} & 12.9 (3.0) & 5, 6 \\ 
4& J0751+1807 & 3.48 & 0.73 & 1.1 & 0.47 & 7, 8 \\ 
5& J1012+5307 & 5.26 & 0.97 & $>$0.77\tablenotemark{d} & 1.3 & 9, 8 \\ 
6& J1024--0719 & 5.16 & $<$0.30\tablenotemark{c} & $<$0.20\tablenotemark{d} & 0.2 & 10, 11, 12 \\ 
7 & J1744--1134 & 4.07 & 0.71 & 0.357\tablenotemark{d} & 0.2 & 10, 13, 12 \\ 
8 & B1937+21 & 1.56 & 10.6 & 3.6 & 3.3 & 11, 14 \\ 
9 & B1957+20 & 1.61 & 1.15 & 2.5 & 0.9 & 11, 15 \\ 
10 & J2124--3358 & 4.93 & 1.30 & 0.25 & 3.6 & 10, 11, 12 \\ 
11 &B1620--26 (M4) & 11.07 & $\sim$0.67\tablenotemark{c} & 1.73\tablenotemark{d} & 0.03 & 16, 17, 18    \\
12 & J1740--5340 (NGC 6397) & 3.60 & 3.90 & 2.55\tablenotemark{d} & 0.24 & 19, 20, 21 \\ 
13 & B1821--24 (M28) & 3.05 & 162 & 5.5\tablenotemark{d} & 4.6 &  22, 23 \\ 
14 & J1911--6000C (NGC 6752) & 5.28 & 0.22 & 4.1\tablenotemark{d} & 0.01 & 24 \\
15 &J2140--2310A (M30) & 11.02 & $<$209\tablenotemark{c} & 9.0\tablenotemark{d} & $\lesssim$0.007 & 25, 26 \\

\enddata

\tablenotetext{a}{For the globular cluster MSPs the host cluster is given in parentheses.}

\tablenotetext{b}{Total flux. For J0437-4715 the value in parentheses gives the non-thermal flux.}

\tablenotetext{c} {Poorly determined acceleration corrections.}

\tablenotetext{d}{Accurate distances/limits. Other estimates are obtained primarily from dispersion measure together with the Cordes \& Lazio (2002) electron density model and are rather uncertain (up to a factor of $\sim$2).}

\tablerefs{(1) Lommen et al. 2000; (2) Becker \& Achenbach 2002; (3) Navarro et al. 1995; (4) Webb, Olive, \& Barret 2004;
(5) van Straten et al. 2002; (6) Zavlin et al. 2002; (7) Lundgren, Zepka, \& Cordes 1995; (8) Webb et al. 2004; (9) Lange et al. 2001; (10) Toscano et al. 1998; (11) Toscano et al. 1999a; (12) Becker \& Tr\"umper 1999; (13) Toscano et al. 1999b; (14) Cusumano et al. 2003; (15) Stappers et al. 2003; (16) Lyne et al. 1988; (17) Richer et al. 1997; (18) Bassa et al. 2004;  (19) D'Amico et al. 2001; (20) Bassa \& Stappers 2004; (21) Grindlay et al. 2002; (22) Cognard \& Backer 2004; (23) Becker et al. 2003; (24) D'Amico et al. 2002; (25) Carreta et al. 2000; (26) Ransom et al. 2004.} 
\end{deluxetable*}

\subsection{Uncertainties in X-ray Luminosity and Pulsar Parameters} 

In comparing theoretical predictions and observed relations it
is necessary to examine and quantify all the effects that act to modify
the observed parameters under consideration.  First, it is important
to emphasize that the X-ray luminosities plotted in Figures 5 and 6
represent observed, not true quantities. This is especially critical
for the non-thermal magnetospheric emission, which is subject to
beaming and, hence, viewing angle effects. Therefore, the measured
value of $L_{X}$ may greatly underestimate the true X-ray luminosity.
As for the thermal emission, a major source of uncertainty is
introduced by the lack of knowledge of the masses and radii of the
sample of MSPs.  These parameters affect $L_{X}$ through the
gravitational redshift and bending of light. In particular, the true
thermal luminosity, as measured at the NS surface, is larger by a
factor of $(1+z_{g})^2$ than the luminosity measured by a distant
observer. In our analysis, we have corrected for this effect by
assuming the canonical values of $M=1.4$ M$_{\odot}$ and
$R=10$ km. However, for a reasonable range of NS masses,
1.2-2.3 M$_{\odot}$, based on the results of \citet{Lyne04} and
\citet{Nice05}, we find that the true $L_{X}$ may differ from the
value used here by up to a factor of $\sim$1.8. In addition, for NS
radii in the range $9-16$ km, the derived $L_X$ may deviate from the
true value by $\lesssim$20\%.

The parameters $\dot{E}$, $B_{\rm surf}$, and $B_{\rm lc}$ also depend
on the MSP mass and radius, as $\dot{E}\propto MR^2$ and $B \propto
\sqrt{MR^2}/R^3$.  For the same range of NS masses ($1.2-2.3$
M$_{\odot}$), the assumption of 1.4 M$_{\odot}$ may result in a
deviation of up to $\sim$60\% for $\dot{E}$ and $\sim$25\% for $B_{\rm
surf}$ and $B_{\rm lc}$, from the true values. This implies that a
spread in masses within the sample of MSPs would tend to increase the
scatter about the derived relations in both coordinates (e.g. $L_X$
and $\dot{E}$) and perhaps result in a skewed trend.  Conversely, for
a sample of MSPs that have very similar masses but are systematically
more or less massive than the canonical 1.4 M$_{\odot}$, the result
will be an offset in the derived intercept, with little effect on the
slope of the log-log relation under consideration (e.g. $\log L_X-\log
\dot{E}$).  As for $R_{\rm NS}$, for the range $9-16$ km we find that
$\dot{E}$, $B_{\rm surf}$, and $B_{\rm lc}$ may differ by up to a
factor of $\sim$2.6. The effect of the NS radius on the scatter in the
trends discussed above is uncertain due to lack of knowledge of the
true NS equation of state (EOS). For instance, in certain theoretical
models of the NS EOS a small change in $M_{\rm NS}$ does not result in
an appreciable change in $R_{\rm NS}$, while for others the opposite
is the case \citep[see e.g.,][for details]{Latt01}.

For the 47 Tuc MSP, the parameters derived from $\dot{P}$ are subject
to an additional uncertainty arising from lack of knowledge of the
magnitude of acceleration of the MSPs by the gravitational potential
of the globular cluster. It is important to note that the
model-dependent values derived for $\dot{P}_i$ are sensitive to the
assumptions made regarding the gravitational potential and gas density
profile of the cluster. For simplicity, in this paper we have derived
$\dot{P}_i$ for each MSP assuming a simple constant gas density
profile.  However, it is likely that in reality the gas distribution
of 47 Tuc is more complex. For instance, the nearly identical
dispersion measures of the MSPs near the cluster center, namely 47 Tuc
F, L, O, and S \citep{Freire03}, may be interpreted as being due to
the presence of a cavity in the plasma around the cluster core carved
out by the winds of these pulsars. Hence, obtaining accurate
values of $\dot{P}_i$ may require implementation of more elaborate
models of the gas distribution, which will be explored in detail
elsewhere.

Another potential source of systematic uncertainty is the presence of
an observational selection effect. It is probable that so far pulsar
observations have sampled only the most luminous and nearby MSPs.  The
existence of X-ray faint MSPs, both in globular clusters and the
field of the Galaxy, would mean that the observed narrow range of
$L_X$ for the MSPs with thermal spectra is simply an artifact of the
limited sample of very low luminosity MSPs.
  
Given the large uncertainties discussed above, any conclusions drawn
for the apparent relations between $L_{X}$ and quantities derived
from the pulsar spin parameters should be taken with much caution, as
it is not possible, at present, to properly account for all effects
that act to modify the observed trends.

\section{DISCUSSION}

\subsection{X-ray Properties of MSPs}

The spectral fits described in \S 3 show that the majority of 47 Tuc
MSPs exhibit emission consistent with a thermal model, with no
indication of any magnetospheric emission.  It is probable that the
MSPs with seemingly purely thermal emission do, in fact, have a
fainter magnetospheric X-ray component ($<$$10^{30}$ ergs s$^{-1}$).
However, it remains unclear what determines the relative strengths of
these two X-ray production mechanisms. Although viewing angle
dependences and beaming of the magnetospheric radiation can explain
why thermal emission dominates in many MSPs, it is more likely that
the relative strength of the two components is determined by the
poorly understood details of the magnetospheric emission and polar cap
heating mechanisms.
 
\citet{Grind02} and \citet{Cheng03} have raised the possibility that
the very narrow range of thermal luminosities and small emission areas
imply the existence of a strong multipole magnetic field near the
surface. However, the small range in $L_{\rm X}$ \textit{is}
consistent with a simple dipole field if one considers general
relativistic effects near the surface.  In particular, gravitational
bending of light causes $\gtrsim$75\% of the NS surface to be visible
at any given time \citep[see e.g.][]{Pech83,Pavlov97,Belo02}. As a
consequence, the degree of modulation in the flux which arises due to
rotation of the NS is greatly reduced.  This, in turn, diminishes the
scatter in $L_{X}$ for the sample of MSPs that would otherwise result
from different orientations of the magnetic and spin axes relative to
the line of sight. Thus, although entirely plausible, the existence of
a multipole magnetic field need not be invoked to account for the
small range in $L_{X}$.

We note that the inferred radii of the thermally emitting areas
obtained in \S 3 are indeed somewhat smaller than the expected size of
a magnetic polar cap for a simple dipole field, $R_{\rm pc}=(2\pi
R_{\rm NS}/cP)^{1/2}R_{\rm NS}$ (e.g., $R_{\rm pc}=2.6$ km for a 3 ms
pulsar).  This likely arises due to the fact that we have fitted most
MSP spectra with a single temperature model, where in reality a
two-temperature model is more appropriate, as suggested by the results
of \S3.  In adittion, due to variation of the projected area of the
polar caps, induced by the rotation of the star, fits to the
time-integrated spectrum result in an underestimate of the effective
emission area. However, even with a two temperature spectrum and a
correction factor for the effective area, in the case of a BB model
the total emitting region is still substantially smaller than the
expected polar cap area.  If the NS surface does, in fact, radiate
like a blackbody, this discrepancy may be indicative of either
non-uniform polar cap heating or deviations from a pure dipole
field. For instance, the polar cap model of Harding \& Muslimov
predicts that the heating rate by positrons should be highest near the
rim of the polar cap \citep[cf. Fig. 7 of][]{Hard02a}, resulting in an
annulus-shaped emitting region with an area significantly smaller than
that of the entire polar cap. \citet{Chen98} have argued that a small
emission area can arise if the MSP is an aligned rotator, resulting in
a so-called ``squeezed'' polar cap. The outer gap model of
\citet{Zhang03} also predicts that the heated area should be
significantly smaller than the whole polar cap due to the presence of
a small scale multipole field near the surface.  Finally, this result
may imply that a blackbody is simply not a valid approximation of the
NS surface. The discrepancy may, thus, be a consequence of the fact
that simplified blackbody models result in higher derived temperatures
and smaller emission areas due to the fact that the X-ray spectra of
light-element atmospheres are harder than blackbody spectra
\citep{Zavlin96}.

The results of this study as well as recent \textit{Chandra} and
\textit{XMM-Newton} observations of nearby field MSPs
\citep{Zavlin02,Beck02,Webb04a,Webb04b} have revealed that there are
no clear systematic differences between the X-ray properties of
globular cluster and field MSPs, which could arise, for instance, due
to different evolutionary paths. For example, the nearby field MSPs
such as J0437--4715 and J0030+0451, are quite similar to the bulk of
47 Tuc MSPs in terms of X-ray luminosity and spectral properties.
Also, B1821--24 in the globular cluster M28, is akin to B1937+21,
characterized by large X-ray and spindown luminosities
($L_{X}>10^{33}$ ergs s$^{-1}$, $\dot{E}>10^{36}$ ergs s$^{-1}$), hard
non-thermal spectra, and very narrow radio and X-ray pulse profiles
\citep{Beck99}.  These similarities imply that if, in fact, multiple
binary exchanges and accretion episodes occur for globular cluster
MSPs, they may not significantly alter the elementary emission
properties of the pulsar.

\subsection{The Eclipsing Binary MSPs}

\citet{Bog05} have shown that the binary MSP 47 Tuc W exhibits
large-amplitude X-ray variability at the binary period. This
modulation can be interpreted as being due to geometric occultations
of an X-ray emitting intrabinary shock by the $\sim$0.15 M$_{\odot}$
main-sequence companion. On the other hand, in the three other systems
in which radio eclipses are observed, namely 47 Tuc J, O, and R, we
detect no such variability. The persistence of X-ray emission during
the radio eclipses confirms that at radio frequencies J, O, and R are
not eclipsed by their respective binary companions but rather by a
tenuous ionized envelope formed due to mass loss by the companion
driven by the pulsar wind, as in the case of PSRs B1957+20 and
J2051--0827 \citep{Arzo94,Stap96}.

The unique X-ray behavior of 47 Tuc W is most certainly the result of
the presence of a main-sequence secondary star instead of the typical
very-low mass ($\sim$0.03 M$_{\odot}$) degenerate companion found in
``black widow''-like systems such as J, O, and R. While the latter
three systems are expected end products of low-mass X-ray binary
evolution, 47 Tuc W is likely the result of an exchange encounter in
which the original very low-mass companion was ejected from the binary
\citep[and references therein]{Freire05}. However, even if the current
companion of MSP W is Roche-lobe filling, the existence of an
intrabinary shock implies that another accretion phase cannot commence
as the energetic pulsar wind is constraining the outflow of mass from
the companion \citep[see][for details]{Bog05}.

\subsection{X-ray versus Radio MSP Properties}

In \S 5 we have found that the $\log L_{X} - \log\dot{E}$ relation for
the thermally emitting MSPs has a slope of $\beta\sim0.2\pm1.1$.
Although such a trend is consistent with the prediction of the polar
cap heating model of \citet{Hard02a}, $L_X\propto\dot{E}^{0.5}$, we
cannot exclude the possibility that the 47 Tuc MSPs follow the linear
$L_{X} - \dot{E}$ relation.

Despite the large uncertainties, for the present sample of MSPs
detected in X-rays we still observe a general trend of decrease of
$L_{X}$ as a function of spin down luminosity and magnetic field
strength. If these parameters are, in turn, determined by the age of
the MSP, it is tempting to speculate that Figures 5 and 6 represent
evolutionary diagrams. Here we consider a possible evolutionary path
of a MSP described within the framework of the \citet{Hard02a}
electron-positron ($e^{\pm}$) pair production model.  For relatively
young and energetic MSPs, such as PSR B1821--24 ($\tau\sim3\times10^7$
yr), $e^{\pm}$ pairs are produced via curvature radiation (CR). These
pairs, in turn, produce the observable X-ray/$\gamma$-ray and radio
non-thermal magnetospheric emission.  Although some polar cap heating
by a backflow of energetic particles may occur in these objects, the
non-thermal component ($\sim$$10^{33}$ ergs s$^{-1}$) is several
orders of magnitude brighter, making it difficult to observe any
emission from the polar caps.  As the MSP ages, it undergoes
rotational spindown and perhaps magnetic field decay, which result in
a decline in the CR pair production rate.  At a certain stage, the
magnetic field becomes too low to permit CR pair production so inverse
Compton scattering (ICS) of soft X-rays by accelerated electrons
becomes an important channel for $e^{\pm}$ pair production. ICS pair
production results in substantial polar cap heating by a backflow of
positrons from the magnetosphere.  Since the polar cap heating rate
has a weaker dependence on the spindown luminosity
($L_h\propto\dot{E}^{1/2}$) than magnetospheric emission
($L_m\propto\dot{E}$), the thermal emission will eventually become
dominant over the magnetospheric emission, as is observed in the
majority of 47 Tuc MSPs and many field MSPs.  At much later times, the
MSP is expected to cross the ICS pair production death line \citep[see
Fig. 1 in][]{Hard02b} and will no longer be observable as a pulsar,
both in the radio and X-ray. In the end, as the polar caps cool, the
MSP will fade away from view at X-ray energies as well.

\section{CONCLUSION}

We have presented a spectral analysis of the X-ray emission for the
complete sample of known MSPs in 47 Tuc. The X-ray spectra of the
majority of MSPs are well described by a thermal model, a BB or a NS
hydrogen atmosphere, with $T_{\rm eff}\sim(1-3)\times10^6$ K, $R_{\rm
eff}\sim0.1-3$ km, and $L_{X}\sim10^{30-31}$ ergs s$^{-1}$ (0.3--8
keV).  The small observed scatter in the thermal $L_{X}$ is consistent
with a simple dipole magnetic field configuration. A two temperature
thermal model, reminiscent of that observed from some nearby field
MSPs, is also consistent with the emission from the bulk of 47 Tuc
MSPs and, in some cases, results in an improved fit. Such a spectrum
could arise due to non-uniform polar cap heating or lateral diffusion
of heat accross the NS surface away from the regions directly heated
by particles from the magnetosphere. The MSPs J, O, and W show a
prominent non-thermal component in their spectra.  At least part of
the non-thermal X-rays are very likely due to interaction of the
relativistic particle wind of the pulsar with matter from the
companion.  All other 47 Tuc MSPs exhibit no evidence of a significant
non-thermal X-ray component.

Our re-examination of the $L_X-\dot{E}$ relation has revealed that
regardless of the emission mechanism responsible for production of the
observed X-rays, the conversion efficiency of $\dot{E}$ to $L_{X}$
appears to be, in all cases, $\sim$$10^{-4}-10^{-3}$.

All of the 47 Tuc MSPs, with the exception of 47 Tuc W, display no
large amplitude variability in their X-ray emission on timescales of
days, weeks, and years. The MSP W exhibits dramatic variations at the
binary period, which can be attributed to the presence of an
intrabinary shock that is eclipsed by the secondary star. The lack of
long term variations in the flux of most MSPs could potentially serve
as a criterion for classification of unidentified soft X-ray sources
\citep{Heinke05}.

Given that most of the $\sim$$100$ MSPs presently known have spin
parameters similar to those of the 47 Tuc MSPs, we expect many of them
to have predominantly thermal X-ray spectra. Moreover, for these MSPs,
gravitational bending of light near the NS surface ensures that at
least one of the antipodal heated polar caps is always observable,
while for a wide range of angles between the magnetic and spin axes
and the line of sight both polar caps are visible simultaneously for a
large portion of the MSP spin period \citep[see e.g.][]{Noll89}. This
effect invariably results in a 100\% duty cycle, even if the
magnetospheric emission beams never sweep across the Earth or if the
radio emission is very faint or heavily absorbed/scattered. Therefore,
many MSPs with heated polar caps may, in principle, be detectable at
X-ray energies even if they cannot be observed as radio pulsars.

Finally, based on our results, we conclude that in terms of the
characteristics of their X-ray emission, there is no clear separation
between MSPs in the field of the Galaxy and in globular clusters.

\acknowledgements 
We would like to thank Peter Edmonds for insightful
discussions, Don Lloyd for use of his NS hydrogen atmosphere model,
and the referee Natalie Webb for many helpful comments. This work was
supported by NASA \textit{Chandra} grants GO2-3059A and G02-3059B. The
research presented here has made use of the NASA Astrophysics Data
System.

\end{document}